\documentclass{elsart1p}
\usepackage{graphicx}
\usepackage{amssymb}
\begin{document}
\begin{frontmatter}
% use the corauthref command within \author for corresponding author
% footnotes;
\title{Standard Model Higgs Boson Searches at the D\O\ Experiment}
\author{Philip Rich}
\address{School of Physics and Astronomy,
The University of Manchester,
Oxford Road,
Manchester,
UK, \\
on behalf of the D\O\ collaboration.
}
\begin{abstract}
We present the latest searches for the Standard Model Higgs boson at a centre-of-mass energy of $\sqrt{s}=$ 1.96 TeV with the D\O\ detector at the Fermilab Tevatron collider. 

%The major contributing process at low mass is associated production ($WH\rightarrow l\nu b\bar{b}$, $ZH\rightarrow\nu\bar{\nu}b\bar{b}$, $ZH\rightarrow llb\bar{b}$ and $WH\rightarrow WWW*$). At high Higgs mass (greater than ~140 GeV) the dominating process is gluon fusion ($gg\rightarrow H\rightarrow WW*$) and is searched for via leptonic decay of the W boson. Analyses presented here use about 3 fb$^{-1}$ of data and a combined limit using all available channels is given.
\end{abstract}
\begin{keyword}
Higgs
\PACS MAN/HEP/2009/6 \sep FERMILAB-CONF-09-014-PPD
\end{keyword}
\end{frontmatter}
%
% main text
\section{The Higgs Boson}
\label{HB}

To explain the origin of mass for the $W$ and $Z$ Bosons in the Standard Model (SM) the Higgs mechanism predicts a breaking of electroweak symmetry. A consequence of this symmetry breaking is a heavy scalar boson whose mass is not predicted by the SM.

Direct searches for the Higgs Boson were performed at the LEP experiments in the process $e^{+}e^{-}\rightarrow ZH$ with a centre of mass energy of 206.6 GeV. A direct mass limit at $m_{H}$ $=>$ 114.4 GeV is set at the 95\% confidence level \cite{LEPlimit}(all limits in this paper will be shown at 95\% CL). This limit is slightly below the maximum available kinematic limit due to a small excess observed in the LEP data.

\begin{figure}[t]
\begin{center}
\includegraphics[width=.40\textwidth]{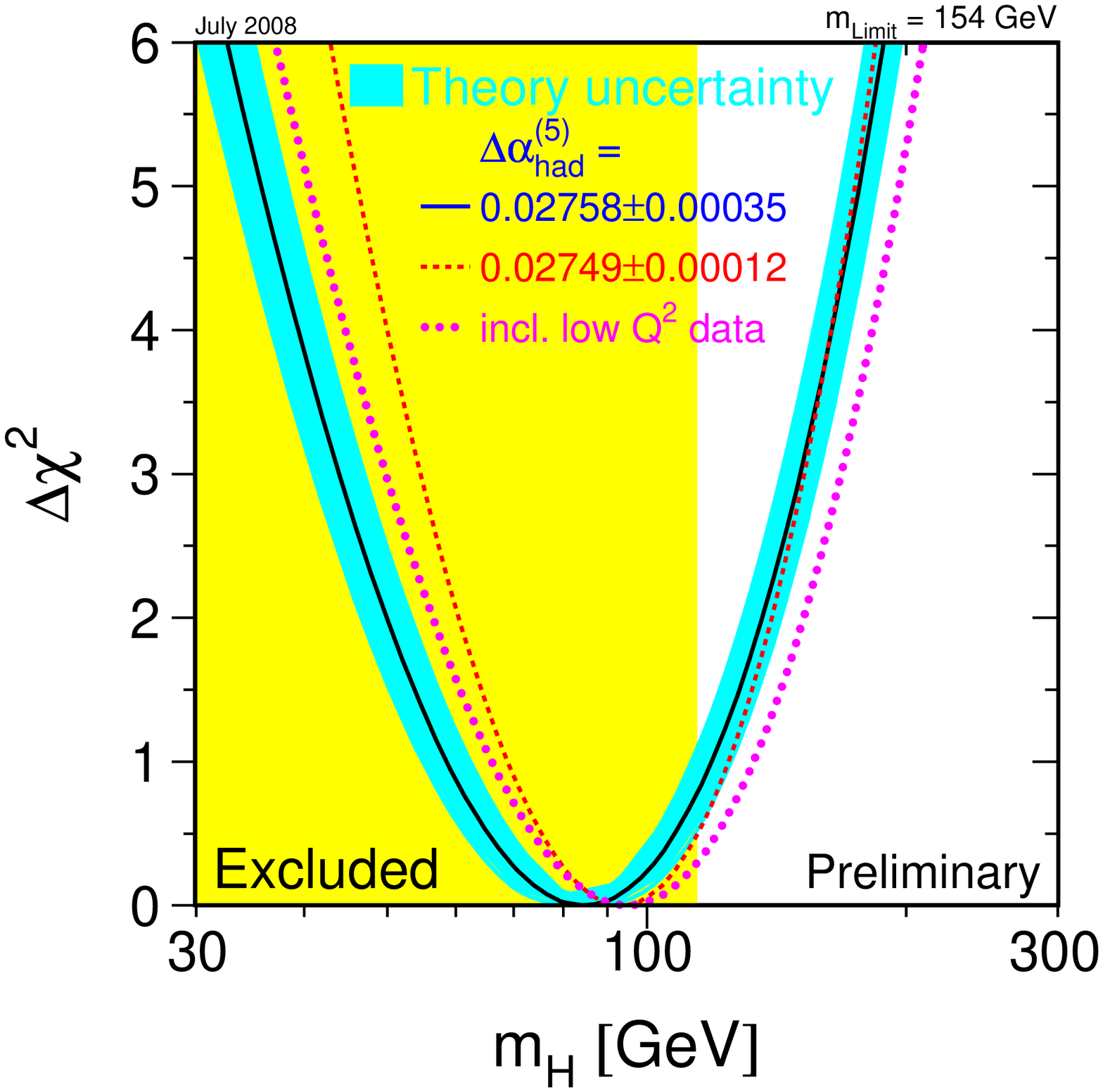}
\includegraphics[width=.40\textwidth]{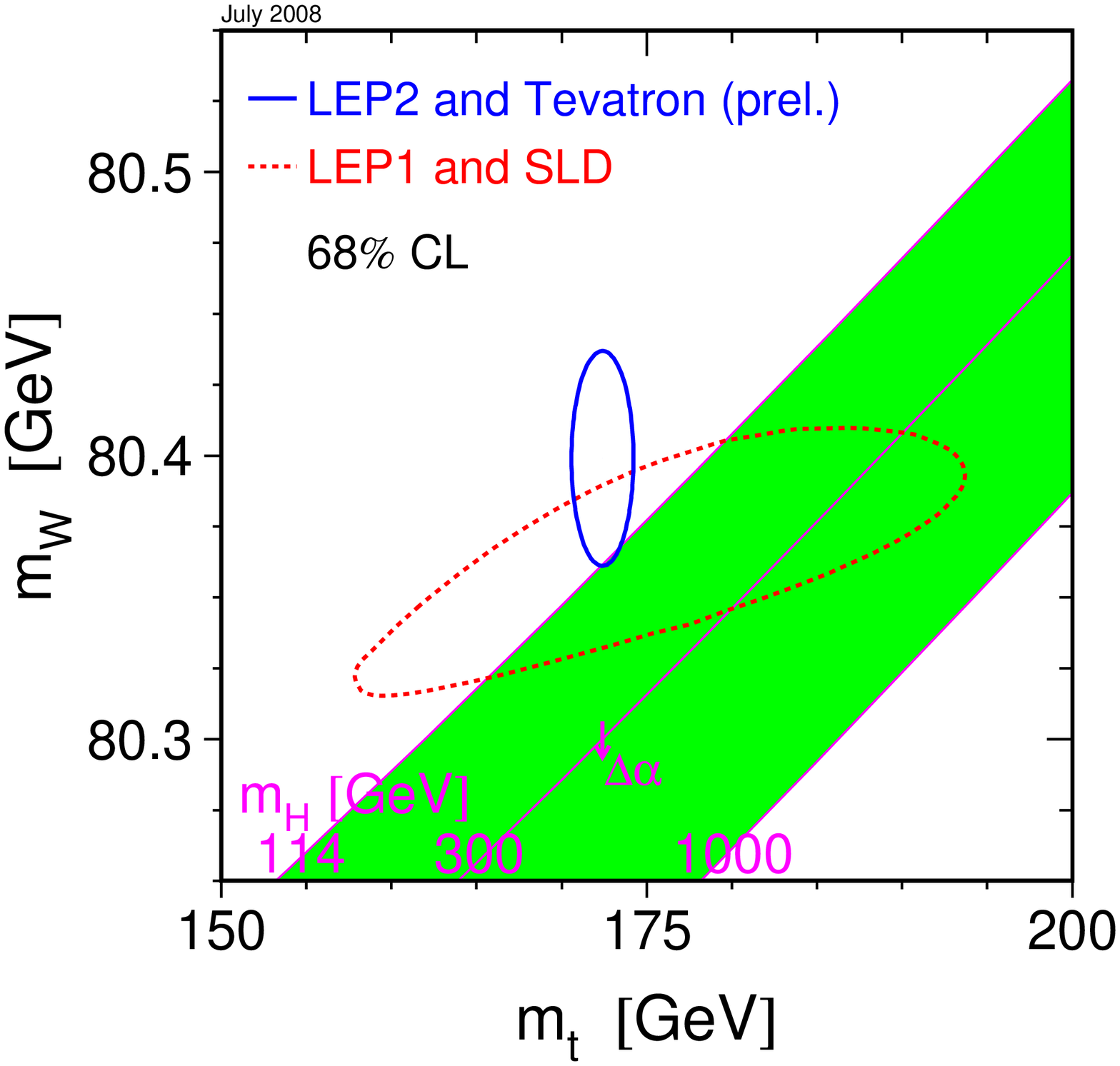}
\end{center}
\caption{Fit for the Higgs Mass from EW data showing the direct search LEP limit and the constraints on the Higgs mass from precision top and W mass measurements \label{fig-BBplot}}
\end{figure}

Indirect limits have been placed on the Higgs boson at the LEP, SLD and Tevatron experiments from electroweak precision measurements \cite{EWconst}. The SM fit best value for the mass of the Higgs boson is $m_{H}$ = 76$^{+36}_{-24}$ GeV \cite{LEPfit}. The upper limit achieved from this fit is $m_{H}
< $ 144 GeV. When the direct mass limit is taken into account this limit is increased to $m_{H} < $ 182 GeV (Figure \ref{fig-BBplot}).

\section{Higgs searches at D\O}
\label{DO}

The fit for the Higgs Mass shown in Figure \ref{fig-BBplot} shows that a light Higgs Boson is favoured which is accessible to the Tevatron experiments. The Tevatron accelerator collides $p\bar{p}$ with a centre of mass energy = 1.96 TeV at an instantaneous luminosity up to around 3$\times$10$^{32}$ cm$^{-2}$s$^{-1}$. At the time of writing, the D\O\ experiment has collected nearly 5 fb$^{-1}$ of data and results shown here include between 1 - 3 fb$^{-1}$ of data analysed. 

The dominant production mechanism for a SM Higgs boson at the Tevatron is via gluon-gluon fusion. An order of magnitude lower is associated production with a $W/Z$ boson. At high masses the main search channel is gluon-gluon fusion with the Higgs boson decaying to two $W$ bosons which subsequently decay leptonically. However, at low Higgs masses ($m_{H} < $ 135 GeV) the Higgs decays into $b\bar{b}$ pairs which are extremely difficult to resolve against the large multijet background at a hadron collider. As such associated production is searched for where events are tagged by the leptonic decay of the $W/Z$ boson.

%\section{$WH\rightarrow l\nu b\bar{b}$}
%\label{WH}

\begin{figure}[t]
\begin{center}
\includegraphics[totalheight=.22\textheight]{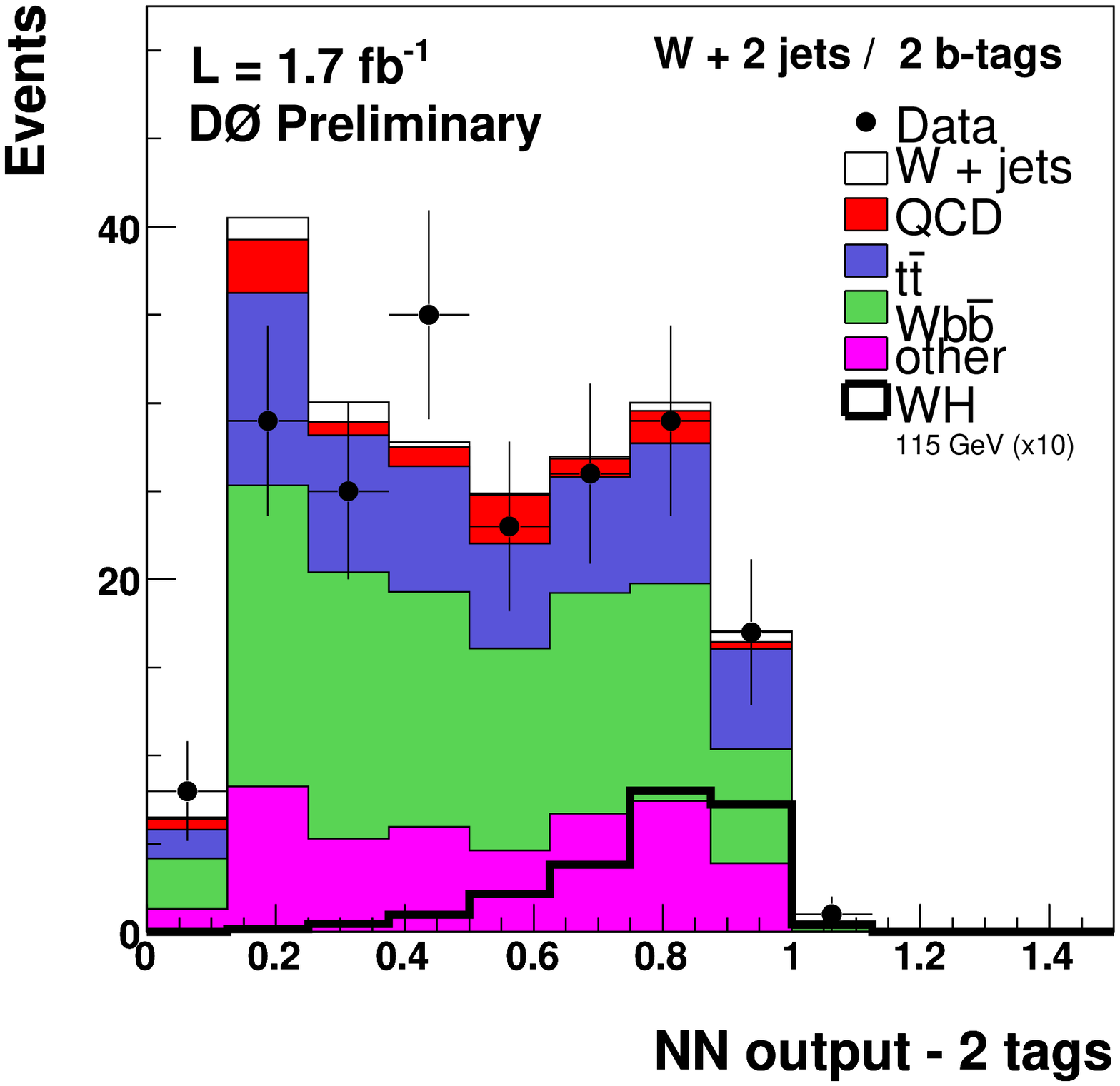}
\includegraphics[totalheight=.22\textheight]{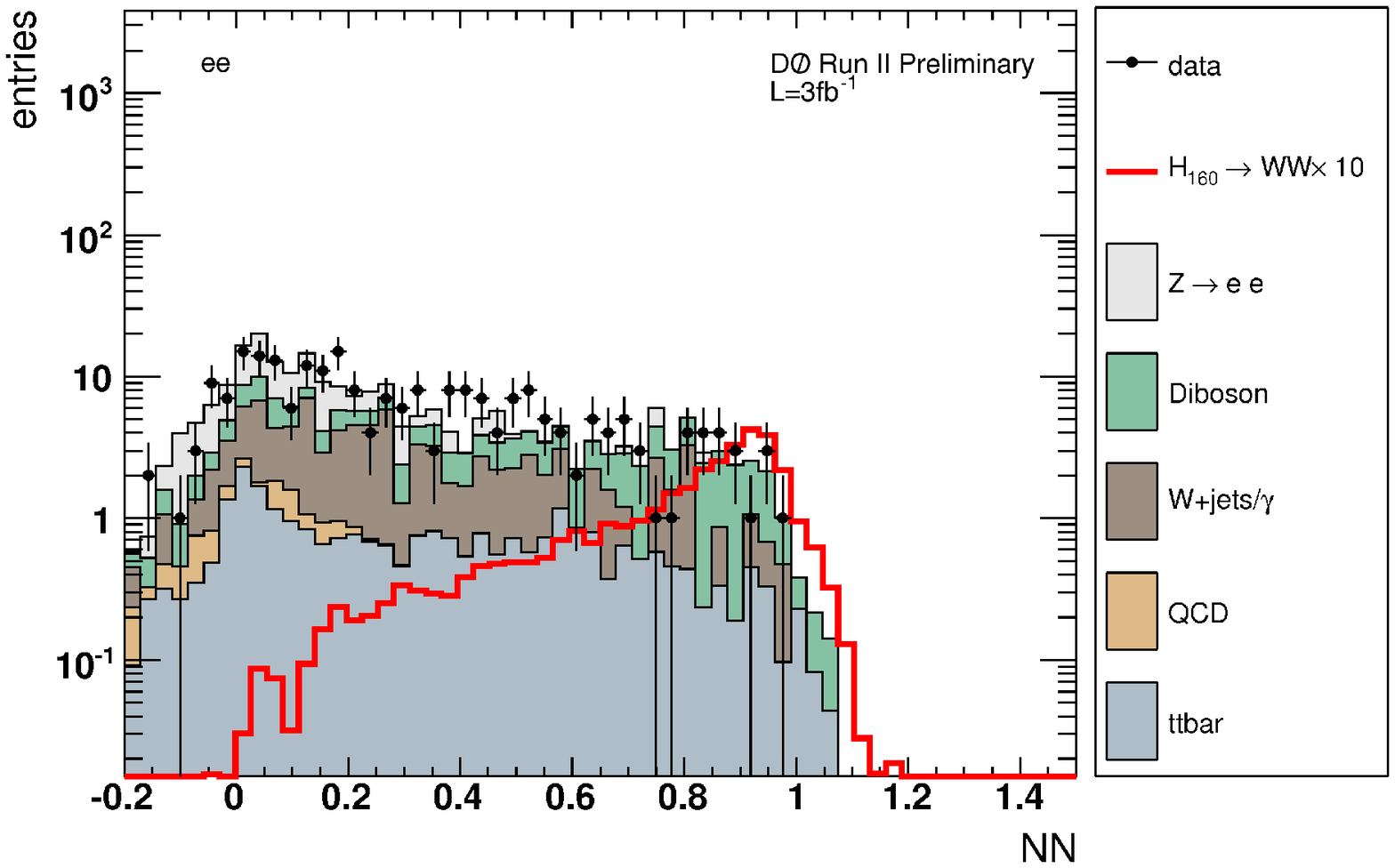}
\end{center}
\caption{Neural Network output distributions for the $WH\rightarrow l\nu b\bar{b}$ and $H\rightarrow WW(\rightarrow l\nu l\nu)$ analyses \label{fig-NNout}}
\end{figure}

One of the most sensitive channels at low mass consists of a final state of two $b$ jets from the Higgs boson and a charged lepton $\ell$ and a neutrino from the W. All three leptonic decays of the W are analysed at D\O\ , however, the most sensitive are the decays to electrons and muons. They will be focused on here using 1.7 fb$^{-1}$ of data. Events are selected with one or two $b$ tagged jets with transverse momentum $p_{T} >$ 20 GeV, an isolated electron/muon with $p_{T} >$ 15 GeV and missing transverse energy $E^{miss}_{T} >$ 20 GeV. The main backgrounds after selection are $W+$jets and $t\bar{t}$ production. To improve separation between the signal and the irreducible background a Neural Network (NN) is trained taking a number of kinematic and topological variables as input. The output of this NN (Figure \ref{fig-NNout}) are used to set limits on Higgs production. The analysis sets a limit on $\sigma_{WH}$ = 10.9x$\sigma_{SM}$ at $m_{H} = 115$ GeV \cite{D0PubRes} (where $\sigma_{SM}$ is the cross section predicted for this process by the Standard Model).

%\section{$H\rightarrow WW(\rightarrow l\nu l\nu)$}
%\label{HWW}

At high mass $H\rightarrow WW$ is searched for in the leptonic final state $\ell\nu \ell\nu$. This analysis uses approximately 3 fb$^{-1}$ of data and searches in three independent final states ($ee,e\mu,\mu\mu$). To select these events two high-$p_{T}$ leptons are required and $E^{miss}_{T} >$ 25 GeV to account for the neutrinos. The main feature of this final state to differentiate from di-boson production is the small angle between the leptons arising from spin-correlations between them due to the spin-0 nature of the Higgs boson. This is in contrast to the back-to-back nature of the leptons from diboson processes. A Neural Network classifier is also used here to improve the discrimination between signal and background and the output (Figure \ref{fig-NNout}) is used to set limits on Higgs production. The analysis sets a limit on $\sigma_{H\rightarrow WW}$ = 1.7x$\sigma_{SM}$ at $m_{H} = 160$ GeV \cite{D0PubRes}.

%\section{Combined Results}
%\label{CR}

\begin{figure}[t]
\begin{center}
\includegraphics[totalheight=.22\textheight]{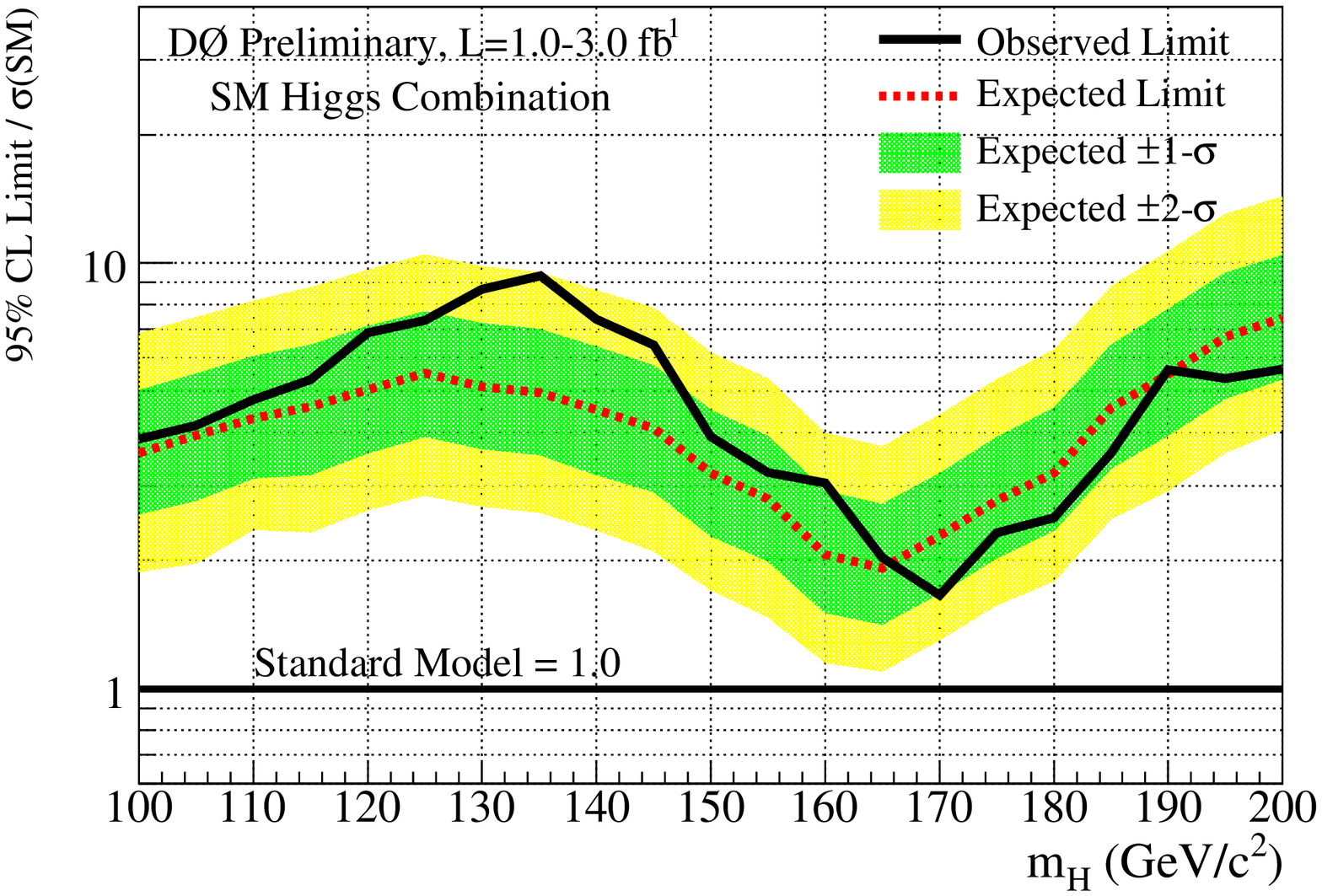}
\includegraphics[totalheight=.22\textheight]{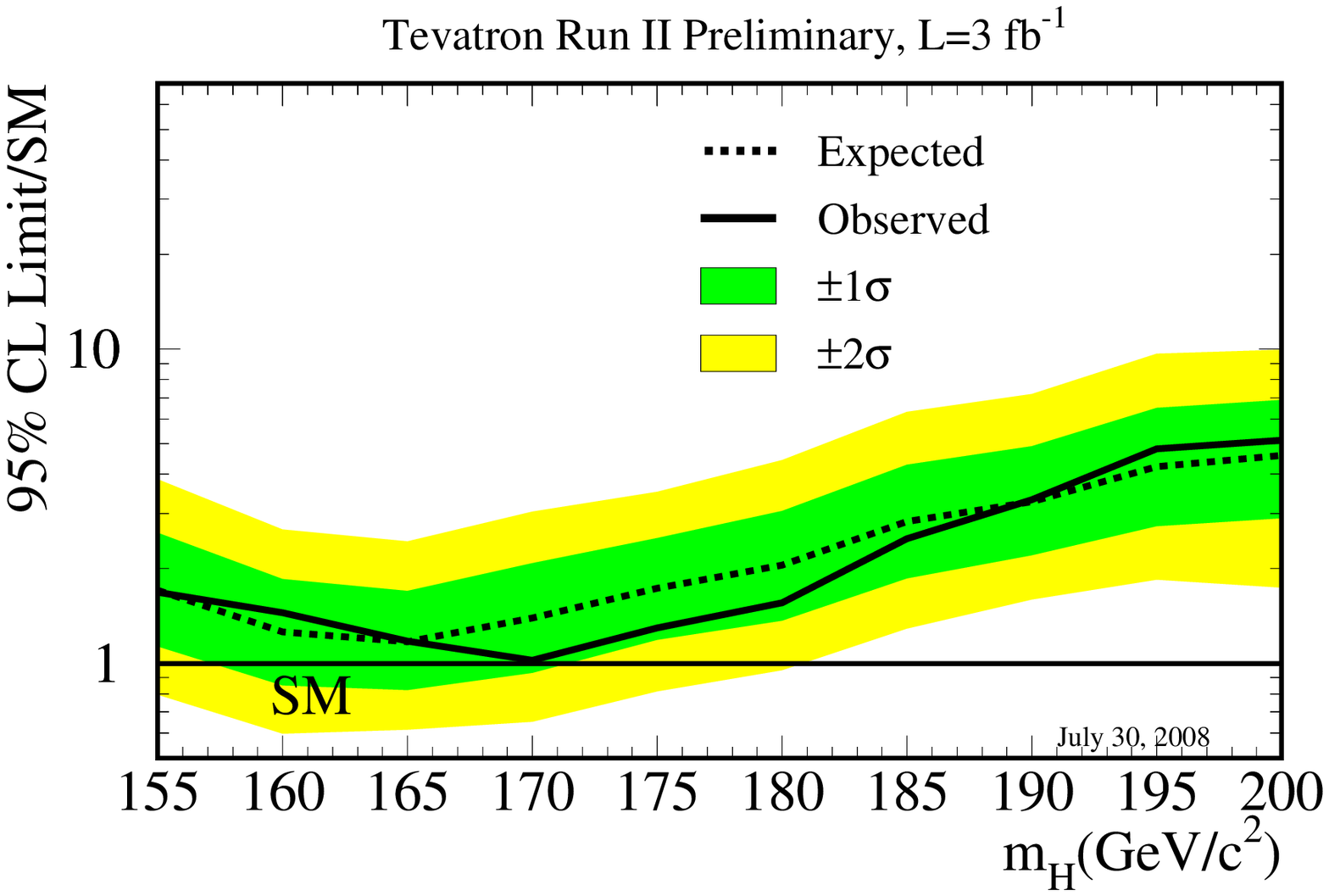}
\end{center}
\caption{Limits on Higgs boson production from D\O\ across the whole mass range and from the Tevatron in the high mass region \label{fig-Limits}}
\end{figure}

Many other channels have been studied in the search for the SM Higgs boson with some more sensitive than others. However, to obtain the best limit on Higgs production at D\O\ a combined limit is set. Systematic uncertainties are taken into account using both Modified Frequentist and Bayesian approaches to ensure the result to does not depend upon the details of the statistical method applied. The results across the full mass range for the D\O\ combination are shown in Figure \ref{fig-Limits}. To further improve the limits on a SM Higgs boson a combination is performed with the CDF experiment (currently only for high mass). The most interesting feature of this combination is that the observed limit crosses one at $m_{H} =$ 170 GeV (Figure \ref{fig-Limits}), resulting in an exclusion of that Higgs boson mass at the 95\% CL \cite{D0PubRes}.

% \appendix
%
% \section{}
% \label{}
%

%
\end{document}